\documentclass
[aps,nofootinbib,preprintnumbers,twocolumn,superscriptaddress,prl]{revtex4}%
\usepackage{amssymb}
\usepackage{epsfig}
\usepackage{bbm}
\usepackage{amsmath}
\usepackage{amsfonts}
\usepackage{graphicx}%
\begin{document}

\title{Observability of Quantum Criticality and a Continuous Supersolid in Atomic Gases}

\author{S. Diehl}

\affiliation{Institute for Theoretical Physics, University of Innsbruck, A-6020 Innsbruck, Austria}
\affiliation{Institute for Quantum Optics and Quantum Information of the Austrian Academy of Sciences, A-6020 Innsbruck, Austria}

\author{M. Baranov}

\affiliation{Institute for Theoretical Physics, University of Innsbruck, A-6020 Innsbruck, Austria}
\affiliation{Institute for Quantum Optics and Quantum Information of the Austrian Academy of Sciences, A-6020 Innsbruck, Austria}
\affiliation{RRC ``Kurchatov Institute'', Kurchatov Square 1, 123182 Moscow, Russia}
\author{A. J. Daley}
\affiliation{Institute for Theoretical Physics, University of Innsbruck, A-6020 Innsbruck, Austria}
\affiliation{Institute for Quantum Optics and Quantum Information of the Austrian Academy of Sciences, A-6020 Innsbruck, Austria}
\author{P. Zoller}
\affiliation{Institute for Theoretical Physics, University of Innsbruck, A-6020 Innsbruck, Austria}
\affiliation{Institute for Quantum Optics and Quantum Information of the Austrian Academy of Sciences, A-6020 Innsbruck, Austria}

\begin{abstract}
We analyze the Bose-Hubbard model with a three-body hardcore constraint
by mapping the system to a theory of two coupled bosonic degrees of freedom. 
We find striking features that could be observable in experiments, including a 
quantum Ising critical point on the transition from atomic to dimer superfluidity 
at unit filling, and a continuous supersolid phase for strongly bound dimers. 

\end{abstract}
\maketitle




Experiments with atomic quantum degenerate gases representing strongly
interacting systems have reached a level of precision where quantitative tests of elaborate many-body theories have become possible
\cite{Esslinger07,GrimmCross07,Ketterle08,Jin08,BlochTroyer08}. In the
interplay between experiment and theory, the challenge is now to identify
realistic models where quantum fluctuations lead to qualitatively new features
beyond mean field in quantum phases and phase transitions. We study below the
Bose-Hubbard model with a three-body constraint, which arises naturally due to a
dynamic suppression of three-body loss of atoms occupying a single lattice
site \cite{Daley09,Roncaglia09}, and can also be engineered via other methods
\cite{Buchler07}. This constraint stabilizes the system when
two-body interactions are attractive, allowing for the formation of dimers --
bound states of two atoms. The phase diagram then contains a dimer
superfluid (DSF) phase connected to a conventional atomic superfluid (ASF). Remarkably,
this simple but realistic model shows several nongeneric features, which are
uniquely tied to the three-body constraint and could be observed with cold
gases: (i) Emergence of an \emph{Ising quantum critical point} (QCP) on the ASF-DSF phase transition line as a function of density -- which
generically is preempted by the Coleman-Weinberg mechanism \cite{Coleman73},
where quantum fluctuations drive the phase transition first order
\cite{Balents97,Radzihovsky04,Vojta00}, with a finite correlation length; and (ii)
A \emph{bicritical point} \cite{FisherNelson74} in the strongly
correlated regime, which is characterized by energetically degenerate orders,
in our case coexistence of superfluidity and a charge-density wave,
representing a ``continuous supersolid" with clear experimental signatures.

Below we describe the constrained model, and then present a new
analytical formalism for a unified treatment of onsite constraints in bosonic
lattice models, based on an exact requantization of the Gutzwiller
mean field theory. This allows for an analytical treatment of the phenomena
arising here. At the end we discuss the experimental signatures of the latter.

\textit{Constrained model} -- We consider the Bose-Hubbard model on a
$d$-dimensional cubic lattice with a three-body onsite hardcore constraint,
\begin{equation}
H\hspace{-0.1cm}=\hspace{-0.1cm}-J\sum_{\langle i,j\rangle}a_{i}^{\dagger
}a_{j}\hspace{-0.01cm}-\hspace{-0.01cm}\mu\sum_{i}\hspace{-0.1cm}\hat{n}%
_{i}+\tfrac{1}{2}U\sum_{i}\hspace{-0.1cm}\hat{n}_{i}(\hat{n}_{i}%
-1),\;a_{i}^{\dag\,3}\hspace{-0.05cm}\equiv\hspace{-0.05cm}0, \label{eq:H}%
\end{equation}
where $\langle i,j\rangle$ denotes summation over nearest neighbors, $J$ is
the hopping matrix element, $\mu$ the chemical potential, and $U$ the onsite
two-body interaction. The three-body constraint stabilizes the attractive
bosonic many-body system with $U<0$, which we focus on here. 

\begin{figure}[b]
\par
\begin{center}
\includegraphics[width=0.9\columnwidth]{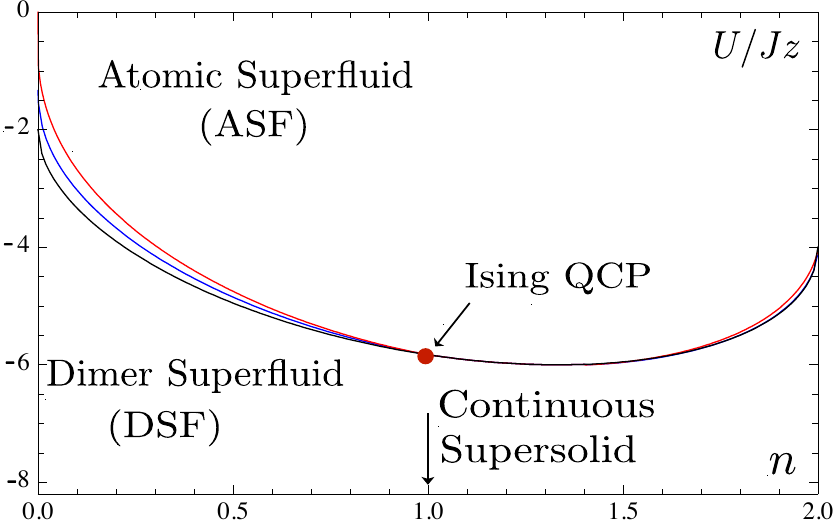}
\end{center}
\par\vspace{-0.5cm}
\caption{Phase diagram for the Bose-Hubbard model with a three-body hard-core constraint, and $U<0$. The black curve
represents the mean field phase border, while red (light gray) and blue (dark gray) curves include
shifts due to quantum fluctuations in $d=2,3$. 
}%
\label{PhaseDiagram}%
\end{figure}

The phase diagram of this model is shown in Fig. \ref{PhaseDiagram}. The
dominant phases are an ASF with order parameters
$\left\langle a_{i}\right\rangle \neq0$ and $\left\langle a_{i}^{2}%
\right\rangle \neq0$, and a DSF with
$\left\langle a_{i}^{2}\right\rangle \neq0$ but $\left\langle a_{i}%
\right\rangle =0$, formed at sufficiently strong interatomic attraction $U$.
In the Gutzwiller mean field approximation \cite{Daley09} these two phases are
separated by a second order phase transition of a special type, the Ising or
ASF-DSF transition \cite{Radzihovsky04}, at which the discrete $Z_{2}$
symmetry of DSF is spontaneously broken. One reason to question the mean field
approach is the presence of two interacting soft modes close to the phase
transition: the noncritical Goldstone mode, related to the $\left\langle
a_{i}^{2}\right\rangle $ order parameter, and the critical Ising mode
signaling the onset of atomic superfluidity with $\left\langle a_{i}%
\right\rangle \neq0$. This motivates the development of a fully quantum
mechanical approach to the constrained Hamiltonian (\ref{eq:H}), with a
\emph{unified} description of interaction effects at various scales.

\emph{Formalism: Mapping the constrained model to a coupled boson theory} --
Because of the constraint, the operators $a_{i},a_{i}^{\dag}$ are no longer
standard bosonic ones and the onsite Hilbert space is reduced to only
three states $|\alpha\rangle$, corresponding to zero
($\alpha=0$), single ($\alpha=1$), and double ($\alpha=2$) occupancy.
Following Altman and Auerbach \cite{Altman02}, for each lattice site $i$ we
introduce three operators $t_{\alpha,i}^{\dag}$ creating these states out of
an auxiliary vacuum state, $|\alpha\rangle_{i}=t_{\alpha,i}^{\dag}%
|\text{vac}\rangle=(\alpha!)^{-1/2}\,\,\left(  a_{i}^{\dag}\right)  ^{\alpha
}|\text{vac}\rangle$. By construction, these operators obey a holonomic
constraint, $\sum_{\alpha}t_{\alpha,i}^{\dag}t_{\alpha,i}=\mathbf{1}$. The
Hamiltonian then reads \vspace{-0.1cm}
\begin{align}
H   =&-J\sum_{\langle i,j\rangle}\big[t_{1,i}^{\dag}t_{0,i}t_{0,j}^{\dag
}t_{1,j}+2t_{2,i}^{\dag}t_{1,i}t_{1,j}^{\dag}t_{2,j}\nonumber%
\\
&   +\sqrt{2}(t_{2,i}^{\dag}t_{1,i}t_{0,j}^{\dag}%
t_{1,j}+t_{1,i}^{\dag}t_{0,i}t_{1,j}^{\dag}t_{2,j})\big]\nonumber\\
& +U\sum_{i}\hat{n}_{2,i}-\mu\sum_{i}\left(  \hat{n}_{1,i}+2\hat{n}%
_{2,i}\right)  ,\label{HtComp}%
\end{align}
where $\hat{n}_{\alpha,i}=t_{\alpha,i}^{\dag}t_{\alpha,i}$. This form is
closely related to resonant Feshbach models used, e.g., to describe the
BCS-BEC crossover in free space \cite{Holland01}: The \textquotedblleft
Feshbach term\textquotedblright\ (second line) allows for interconversion of a
\textquotedblleft dimer\textquotedblright\ ($t_{2,i}$) into two
\textquotedblleft atoms\textquotedblright\ ($t_{1,i}$) on nearby sites, and
the detuning (first term in the last line), gives the energy difference of
atoms and dimers. Here the role of the detuning is played by the onsite
interaction $U$, in contrast to $\sim1/U$ in the resonant models.

Using the constraint, we can eliminate one of the
operators $t_{\alpha,i}$ in Eq. (\ref{HtComp}), say $t_{0,i}$, as $t_{0,i}\rightarrow\sqrt{X_{i}}$
(and $t_{0,i}^{\dag}\rightarrow\sqrt{X_{i}}$), where $X_{i}=1-\hat{n}%
_{1,i}-\hat{n}_{2,i}$. Noting that $X_{i}^{2}=X_{i}$ on the constrained space, we replace $\sqrt{X_{i}}$ with
$X_{i}$, yielding a \emph{polynomial} Hamiltonian. The remaining
operators $t_{1},t_{2}$ can now be interpreted as standard bosonic ones. To demonstrate this, one divides the standard bosonic Hilbert space into a physical ($\mathcal{P}_{i}$) and an unphysical
($\mathcal{U}_{i}$) subspace, $\mathcal{H}_{i}=\mathcal{P}_{i}\oplus
\mathcal{U}_{i}$, where the physical one consists of states obeying the 3-body
constraint. We see that $H$ \emph{does not couple} the physical $\mathcal{P}%
=\prod\mathcal{P}_{i}$ and unphysical $\mathcal{U}=\prod\mathcal{U}_{i}$ subspaces.

The distinction between the contributions from physical and unphysical spaces
can most conveniently be achieved by using the quantum effective action
$\Gamma\lbrack t_{1},t_{2}]$ \cite{AmitBook}, which is a functional on
classical fields and provides all one-particle irreducible correlation
functions as coefficients of an expansion in powers of $t_{1}$ and $t_{2}$.
Because $\Gamma$ is formulated in terms of physical quantities, we can restrict
its general form to a polynomial obeying the three-body constraint. The form of
the effective action thus is restricted by the constraint, in addition to the
symmetries of the microscopic theory.

To apply the above construction to a many-body system, one has to choose the
proper ground state and the corresponding operators. Following Ref.
\cite{Altman07}, we introduce new operators $b_{\alpha,i}=(R_{i})_{\alpha
\beta}t_{\beta,i}$ ($\alpha,\beta=0,1,2$), with a unitary transformation $R$.
The parameters of $R$ are such that $b_{0,i}$ creates the mean field vacuum,
and $b_{1,i}$ and $b_{2,i}$ correspond to fluctuations on top of this state,
with vanishing expectation values (see \cite{Diehl09}). The DSF ground state,
for example, corresponds to: $b_{0,i}=\cos(\theta/2)t_{0,i}+\sin(\theta
/2)\exp(-i\phi)t_{2,i}$, $b_{2,i}=\cos(\theta/2)t_{2,i}-\sin(\theta
/2)\exp(i\phi)t_{0,i}$, and $b_{1,i}=t_{1,i}$, where $\phi$ is an arbitrary
phase and the angle $\theta\in\lbrack0,\pi]$ is such that $2\sin^{2}%
(\theta/2)=n$, the density of atoms (on the mean field level for simplicity). The operators $b_{\alpha,i}$ are subject
to the same constraint, $\sum_{\alpha}b_{\alpha,i}^{\dag}b_{\alpha
,i}=\mathbf{1}$, and we can eliminate $b_{0,i}^{(\dag)}$ as described above.
This results in a \emph{polynomial} Hamiltonian for the remaining operators
$b_{1,i}$ and $b_{2,i}$, where the operator independent part precisely
reproduces the Gutzwiller mean field energy.

\emph{Application of the formalism} -- The limit $n\to0$ gives a stringent check of our formalism, where we recover the 
nonperturbative Schr\"odinger equation for the dimer bound state formation (see \cite{Diehl09} for details). We now apply our method in the many-body context:


\emph{ (i) Ising quantum critical point.} The polynomial Hamiltonian describes
atomic ($b_{1,i}$) and dimer ($b_{2,i}$) fluctuations around the spatially
uniform DSF state. After taking the long wavelength continuum limit, one can
easily see that there are two soft modes in the vicinity of the ASF-DSF
transition: the noncritical Goldstone mode $\pi\sim\mathrm{\operatorname{Im}%
}(b_{2})$ corresponding to the $U(1)/Z_{2}$ gauge symmetry broken by the
presence of the dimer condensate, and the critical atomic Ising mode
$\varphi\sim\mathrm{Re}(b_{1})$ signaling the appearance of an atomic
condensate. The other two modes remain massive and do not affect the physics
of the phase transition. Integrating out the latter we obtain an effective low
energy action for the soft modes \cite{Diehl09}:\vspace
{-0.2cm}
\begin{align}
S_{\mathrm{eff}}[\varphi,\pi]   =&\int_{x}\big\{\tfrac{1}{2}\varphi
(-Z_{\varphi}\partial_{\tau}^{2}-\xi_{+}^{2}\Delta+m_{+}^{2})\varphi
+\lambda\varphi^{4}\nonumber\\
&  +\tfrac{1}{2}\pi(-Z\partial_{\tau}^{2}-\xi^{2}\Delta)\pi+\mathrm{i}%
\kappa\varphi^{2}\partial_{\tau}\pi\big\}.\label{Seff}%
\end{align}
It describes phonons $\pi$ in the dimer superfluid coupled to a real scalar
Ising field $\varphi$, in turn represented by an action of the Ginzburg-Landau
type with the \textquotedblleft mass" parameter $m_{+}^{2}$ crossing zero at
the ASF-DSF transition. The coupling $\kappa$ comes from the cubic coupling
$-\sqrt{2}J\cos(\theta)b_{2,i}^{\dag}b_{1,i}b_{1,j}+h.c.$ which originated from the
\textquotedblleft Feshbach term\textquotedblright\ in the original Hamiltonian
Eq. \eqref{HtComp}, such that $\kappa\sim\cos(\theta)\approx1-n$. This cubic
coupling of Goldstone to Ising mode with linear time derivative has the same
degree of relevance as the Ising coupling $\lambda$, leading to the
Coleman-Weinberg (CW) phenomenon \cite{Balents97}: The renormalized value of
the Ising coupling $\lambda$ reaches zero at some \emph{finite} scale $\xi$ at
which $m_{+}^{2}$ is still positive. As a result, terms with higher powers of
$\varphi$, which are generated by fluctuations, become important. These
self-interaction terms provide a new minimum with $\left\langle \varphi
\right\rangle \neq0$, which is reached via a first order phase transition with
finite correlation length $\xi$. Therefore, the ASF-DSF phase transition in
our model is actually first order, contrary to the predictions of the mean
field approach.

The coupling via a temporal derivative and, therefore, the CW phenomenon, is
rather generic in nonrelativistic systems, in which an Ising field emerges as
an effective low energy degree of freedom
\cite{Radzihovsky04,Vojta00,Balents97}. In our case, however, the coupling
$\kappa$ vanishes for $n\rightarrow1$. The existence of such a decoupling
point can be proven assuming a continuous, monotonic behavior of a particular
compressibility, the $\mu$ -derivative of the dimer mass term $K=-d m_{d}%
^{2}/d\mu|_{n}$ within the DSF phase: it then must have a unique zero
crossing, because it is $>(<)0$ for $n=0(2)$ \cite{Diehl09}. This argument is
tied to the existence of a maximum filling, and thus to the three-body
constraint. Using the Ward identities resulting from a temporally local gauge
invariance $b_{\alpha}\rightarrow\exp(\mathrm{i}\alpha\lambda(t))b_{\alpha}$,
$\alpha=1,2$, and $\mu\rightarrow\mu+\mathrm{i}\lambda(t)$ \cite{SachdevBook},
we see that $\kappa\propto K$. As a result, the first order $Z_{2}$ transition
terminates into a true $d+1$ dimensional Ising quantum critical point in the
vicinity of unit filling.

Fluctuations also shift the mean field phase boundary, cf. Fig.
\ref{PhaseDiagram}. The shift is only pronounced for $n\ll1$, the reason being
that dimer formation and atom criticality approach each other for $n\to0$ \cite{Diehl09}.

\emph{ (ii) Continuous supersolid.} Another peculiar consequence of the three-body
constraint occurs for dimers in the strong coupling limit,
where single particle excitations are strongly gapped ($\sim|U|/2$) and can be
integrated out perturbatively. Taking the dominant nearest
neighbor hopping $t$ and interaction $v$ into account, the effective lattice theory for hardcore bosons (dimers or di-holes)
can be conveniently rewritten as an antiferromagnetic Heisenberg spin
Hamiltonian (see, e.g., \cite{AuerbachBook}), 
\begin{equation}
H_{\mathrm{AF}}=2t\sum_{\langle i,j\rangle}\left[  s_{x,i}s_{x,j}%
\hspace{-0.05cm}+\hspace{-0.05cm}s_{y,i}s_{y,j}\hspace{-0.05cm}+\hspace
{-0.05cm}\lambda s_{z,i}s_{z,j}\right] \label{HAF}  \hspace{-0.05cm}%
\end{equation}
restricted to a subspace with a fixed projection of the total spin on the
$z$-axis, $S_{z}=\sum_{i}s_{z,i}=L(n_{d}-1/2)$, where $L$ is the total number
of the lattice sites, and $n_{d}=n/2$ the dimer filling. The anisotropy
parameter $\lambda$ is the ratio of the interaction and hopping,
$\lambda=v/2t$. In the leading second order perturbation theory, we find
$t=v/2=2J^{2}/|U|$, such that $\lambda=1$ and the Hamiltonian (\ref{HAF}) is
$SO(3)$ -invariant, corresponding to a symmetry enhancement compared to the
conventional $SO(2)\simeq U(1)$ phase symmetry for bosons. It parallels a
similar effect for attractive lattice fermions \cite{Zhang90}, and is a
peculiar feature of the three-body hardcore constraint -- if virtual triple
and higher occupancies are allowed, one finds $\lambda=4$
\cite{Fleischhauer06}. The symmetry enhancement is operative for exactly half
filling of dimers, $n_{d}=1/2$, where $S_{z}=0$, while for other dimer
fillings $S_{z}\neq0$ and the symmetry is reduced to $U(1)$. In
the first case, however, the ground state of the Hamiltonian (\ref{HAF}) is
the antiferromagnetic state parametrized by the direction of the N\'{e}el
order parameter on the three-dimensional Bloch sphere. Generically, the N\'{e}el
order parameter has components both in the $xy$ -plane and along the $z$ -axis.
This means that the ground state of bosons has both DSF and charge-density
wave (checkerboard-like, CDW) orders, i.e. is a supersolid. The specific
feature, however, is that the $SO(3)$ symmetry admits a continuous change in the ratio between the two
order parameters without changing the energy, and this state may thus be
called a continuous supersolid. A particular choice of the order
parameters depends on the way the system is prepared and on the boundary
conditions. This is in contrast to other occurrences of supersolidity
in bosonic systems \cite{SuSo}.

The spontaneously broken $SO(3)$ symmetry in the continuous supersolid
provides us with \emph{two} massless Goldstone modes. A spin wave analysis
yields their dispersion
\[
\omega(\mathbf{q})=tz\left[  \widetilde{\epsilon}_{\mathbf{q}}(1+\lambda
-\lambda\widetilde{\epsilon}_{\mathbf{q}}) \right]^{1/2}
\]
with $\tilde{\epsilon}_{\mathbf{q}}=1/d\sum_{\lambda}(1-\cos\mathbf{qe}%
_{\lambda})$, $\lambda\leq1$. For $\lambda=1$ the
second Goldstone mode emerges at the edge of the Brillouin zone,
adding to the one at zero momentum. In the next (fourth) order of perturbation
theory, we find \cite{Diehl09} $\lambda=1-8(z-1)(J/\left\vert U\right\vert
)^{2}<1$, and the DSF ground state is slightly favored over CDW order due to weak
explicit breaking of $SO(3)$. Still, the proximity to the continuous
supersolid manifests itself in a weakly gapped ($\Delta\sim tz(1-\lambda)$)
collective mode at the edge of the Brillouin zone, which may be probed
experimentally, see below.

\emph{Experimental signatures} -- We now discuss in detail the experimental
signatures of the critical behavior and continuous supersolid. Above we
showed that the ASF-DSF phase transition (see Fig.~1) terminates in an Ising
quantum critical point at $n=1$. The phase transition is thus second order at
$n=1$, with a diverging correlation length at the transition point. Away from $n=1$ the transition is first order, with a finite
correlation length estimated as 
$\xi/a\sim\kappa^{-6}\sim|1-n|^{-6}$ ($a$ the lattice spacing) using the renormalization group flow of \cite{Balents97}. As typical of
the radiatively induced first order transitions, the near-critical domain is
large, with a correlation length exceeding the typical extent of optical
lattices of 50 to 100 sites in a region $1/2\lesssim n\lesssim3/2$. This
behavior can be measured directly in experiments probing the correlation
length, as done in Ref.~\cite{Esslinger07}. Alternatively, critical
opalescence via damping of collective oscillations \cite{JinMewes96} could
be probed.

\begin{figure}[t]
\par
\begin{center}
\includegraphics[width=1\columnwidth]{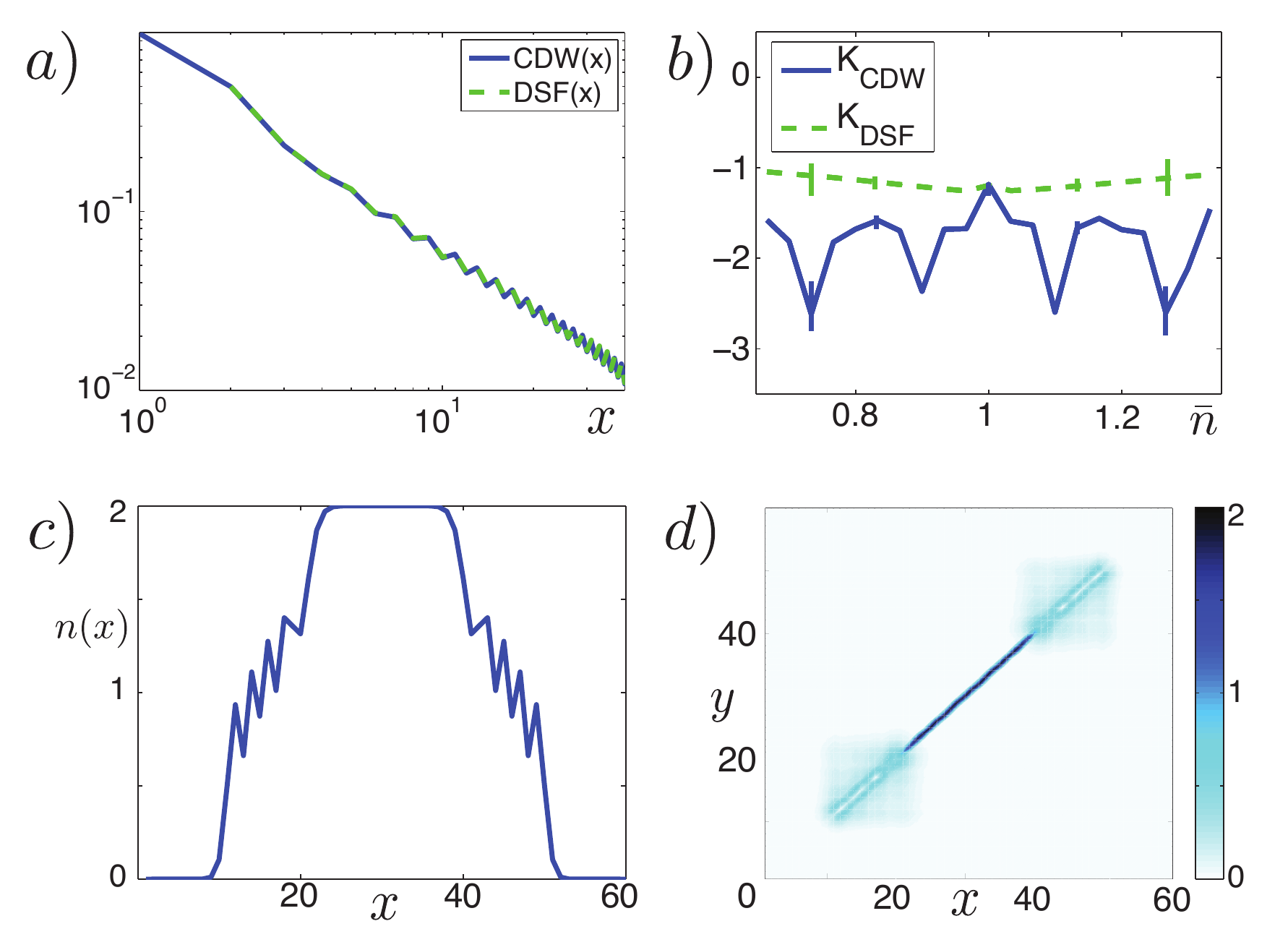}
\end{center}
\par\vspace{-0.7cm}
\caption{Ground state computations on 60 lattice sites with $U/J=-20$ using
t-DMRG. (a) Correlation functions characterizing the CDW and DSF orders for
open boundary conditions at unit filling on a log-log scale as a function of
distance $x$. (b) Fitted algebraic decay exponents $K_{\text{DSF}}$ and
$K_{\text{CDW}}$ for varying $n$ (errorbars show estimates of the fitting
error). (c) Density $n(x)$ for a system with a harmonic trapping potential
$V(x)/J=(x-30.5)^{2}/900, N=60$ particles. (d) Shaded plot of the DSF
correlation function $\langle b_{x}^{\dag}b_{x}^{\dag}b_{y} b_{y}\rangle$ with
interpolated shading, indicating substantial DSF order where $n\sim1$ . }
\end{figure}

The continuous supersolid appearing at strong attraction and unit filling can
be detected by measurement of the coexisting spatial order and dimer
superfluid correlations. The spatial structure could be detected via noise
correlation measurements \cite{Blochnoise}, and the dimer superfluid
correlation functions in the momentum distribution of dimers, which could be
imaged, e.g., by associating atoms to molecules, and measuring their momentum distribution. Strong evidence would also
be obtained by probing collective modes that appear at the edge of the
Brillouin zone. It is also possible to stabilize the CDW by ramping a weak
($\sim\Delta\ll tz$) superlattice, acting as a staggered external field that
rotates the N\'eel order parameter from the $xy$ -plane to the $z$ -axis.

We further elaborate on experimental signatures using exact numerical
time-dependent density matrix renormalization group (t-DMRG) techniques
\cite{tdmrg} in 1D in the ground state for realistic experimental size scales
and parameters:
Fig.~2a shows the density-density correlation function $\langle n_{i} n_{i+x}
\rangle-\langle n_{i} \rangle\langle n_{i+x} \rangle$ characterizing CDW
order, and the DSF correlator $\langle b_{i}^{\dag}b_{i}^{\dag}b_{i+x} b_{i+x}
\rangle$ as a function of distance $x$. At unit filling, both decay
algebraically and are essentially equal, indicating coexistence both orders.
In Fig.~2b we show the result of fitting an algebraic decay
$x^{K_{\text{CDW,DSF}}}$ to the correlation functions. For $n =1$ these are
equal, but away from unit filling, the DSF correlations decay more slowly, so
that DSF order dominates CDW \footnote{The large fluctuations in
$K_{\text{CDW}}$ with varying $n$ are due to the interplay between filling
fraction and CDW order.}. In experiments in a harmonic trap, where the filling
factor varies across the system, this gives rise to further signatures. In
Fig.~2c, we plot the density in a trap, showing that a region exists near unit
filling where oscillations in the density are present, characteristic of the
appearance of CDW order. Fig.~2d shows that in the same region the DSF
correlations are significant, whereas in the center of the trap, a
constraint-induced insulating phase with $n=2$ appears. For more details see
\cite{Diehl09}.

\emph{Conclusion} -- We have demonstrated that an atomic Bose gas in an
optical lattice with three-body onsite constraint provides a realization of
such fundamental physical concepts as the Coleman-Weinberg phenomenon of
radiative mass generation and Ising quantum criticality. In addition, the
ground state at unit filling in the strongly correlated limit is an example of
a continuous supersolid - a supersolid with a tunable ratio between the
superfluid and the charge-density wave order parameters.

\emph{Acknowledgment} -- We thank E. Altman, H. P. B\"{u}chler, M.
Fleischhauer, M. Greiter, A. Muramatsu, S. Sachdev, L. Radzihovsky and J.
Taylor for discussions. This work was supported by the Austrian Science
Foundation through SFB FOQUS, by RFBR, and by the EU IP SCALA.

\end{document}